# Influence by zirconium plastic deformation at temperature of 4.2 K on zirconium crystal lattice structure and magnitude of superconducting transition temperature $T_c$

V. K. Aksenov, B. G. Lazarev, O. P. Ledenyov, V. I. Sokolenko, Ya. D. Starodubov and V. P. Fursa

*National Scientific Centre Kharkov Institute of Physics and Technology,
Academicheskaya 1, Kharkov 61108, Ukraine.*

We have researched the effect of the zirconium deformation by the extension at the temperature of *4.2 K*, with the subsequent heating up to the room temperature of *300 K*, on both the zirconium crystal grating structure and the magnitude of superconducting transition temperature $T_c$, using the zirconium samples, synthesized by the method of the electron beam melting. In our opinion, a registered substantial increase of the critical temperature $T_c$ (by *20%*) is a result of both the superconductivity enhancement in the plastic deformation regions at the twin boundaries of the zirconium crystal grating and the effective change of the inter-electron attraction constant magnitude.

## Introduction

Up to the present date, the influence by the crystal lattice defects on the superconducting transition temperature $T_c$ has been established for a wide circle of superconductors. The researches in this field were stimulated by the *Shal'nikov* [1] research results on the metal films, deposited on the cold substrates, and by *Khotkevich* [2] research results on the bulk superconductors, strained at the low temperatures. However, most of the earlier research results did not include the microstructure studies.

Recently, the effects by the low temperature plastic deformation and by the defects, originated during the straining, on the superconducting transition temperature $T_c$ of the vanadium and niobium (the transition metals with the *bcc*), have been researched [3-6]. It was shown that the transition temperature $T_c$ increases substantially. Similar increase in the magnitude of the critical temperature $T_c$ as a result of the low temperature deformation was also observed in the non-transition metals with the tetragonal (*In* and *Sn* [2, 7, 8]) and hexagonal (*Tl* [2]) crystal lattices. We are interested in the research on the influence by the low temperature straining of superconductor on the critical temperature $T_c$ of superconductor with the *hcp* crystal lattice at the specified conditions, aiming to get a better understanding of the phenomenon in the general case of the metallic superconductors with the different crystal lattice types.

## Synthesis of Samples for Experimental Measurements

The poly-crystal electron-beam-melted zirconium of purity of *99.95 wt. %* of high plasticity was a main object of our research conducted at the low temperatures. As per our knowledge, the experimental researches on this type of zirconium have not yet been carried out, which is likely due to the difficulties that are associated with the low superconducting transition temperature in the initial state, $T_{c0} = 0.5\ K$. In our case, the roll-treated ingot was sliced on a spark machine to obtain the zirconium samples, shaped as the double-ended blades with working part of *15 mm* and *0.25 x 0.40 $mm^2$* in the cross section. After the annealing in the vacuum of *1.3 x $10^{-4}$ Pa* at the temperature of *800 °C* for the time period of one hour, the mean grain dimension was about *50 μm*.

## Experimental Measurements Results

The straining of zirconium by the extension (up to *8 %*) was performed at the temperature of *4.2 K* at the rate of *2·$10^{-3}\ sec^{-1}$* on a setup, described in [9]. After the warm up to the room temperature and maintenance at this temperature for the time of *24* hours, we researched the defects structure of strained zirconium by the means of the transmission electron microscope as well as measured the critical temperature $T_c$ of strained zirconium during its transition from the normal state to the superconducting state. The precise measurements of critical temperature $T_c$ were conducted with the use of an experimental setup, based on the liquid *Helium three* (*$^3He$*) cryostat. The temperatures from *1 K* down to *0.35 K* were produced by the pumping off vapors above the liquid *$^3He$*. The researched sample was placed in the liquid *$^3He$*, which ensured the good thermal contact and stabilization. The heat power of ~ *$10^{-6}$ W* was released in the sample during the measurements. The estimated *Kapitsa* jump was such that it could be neglected in this experimental research.

The sample's temperatures were found from the $^3He$ vapor pressure measurements by the means of the *McLeod* manometer. The accuracies of determinations of both the absolute temperatures and the relative temperatures were $5·10^{-3}$ *K* and $5·10^{-4}$ *K* respectively. The superconducting transition was determined, using the four-probe measurement method. In order to minimize a possible influence by the *Earth's* magnetic field, it was compensated by the two pairs of the *Helmholtz* coils to the level below $10^{-2}$. The current was generated by the current source with the long time stability of not worse than $10^{-4}$. The magnitude of measured current was $5·10^{-2}$ *A*. The increase of current did not result in the temperature changes in the cryostat and did not make a significant shift of superconducting transition temperature. The recording of the *Normal metal – Superconductor* (*N-S*) transition was performed by the nano-voltmeter *F-118*. Table 1 has the data on the magnitudes of measured changes of zirconium electrical resistance ratio $R_{293\ K} / R_{1\ K}$, the values of the zirconium normal state temperature $T_N$ at the *N-S* transition, the values of the zirconium critical temperature $T_C$ at the *N-S* transition, and the range of temperatures between the initial state of zirconium and the final state of zirconium after its low temperature plastic deformation at the temperature of *4.2 K* and its annealing at the temperature of *300 K* at the *N-S* transition.

| Zirconium States | $\frac{R_{293K}}{R_{1K}}$ | $T_N$, K | $T_S$, K | $\Delta T$, K |
|---|---|---|---|---|
| Initial State | 110 | 0.470±0.005 | 0.451±0.005 | 0.020 |
| Final State | 63 | 0.585±0.005 | 0.530±0.005 | 0.055 |

***Tab. 1.** Experimental measurements data:*
*1) magnitudes of zirconium electrical resistance ratio $R_{293\ K} / R_{1\ K}$,*
*2) zirconium normal state temperature $T_N$ at N-S transition,*
*3) zirconium superconducting transition temperature $T_C$ at N-S transition,*
*4) range of temperatures $\Delta T$ between temperature $T_N$ in initial state and temperature $T_S$ in final state at N-S transition of zirconium after its low temperature plastic deformation at temperature of 4.2 K and annealing at temperature of 300 K.*

As it follows from the analysis of obtained experimental results, just as in the cases of the metals with the *bcc* lattice (*Nb, V*) and the tetragonal lattice (*In, Sn*), the low temperature plastic deformation of zirconium with the *hcp* crystal lattice, even with the subsequent heating up to the temperature of *300 K*, does not result in a noticeable increase of the critical temperature $T_C$. In this case, the absolute increase of critical temperature $T_C$ of zirconium was equal to ~*0.1 K*, which is as big as *20 %* of the initial value of critical temperature $T_C$. This value is bigger by an order of magnitude than the highest increase of the critical temperature $T_C$ in the niobium and vanadium, where the ratio $\Delta T_C/T_{C0}$ after the heating up of samples amounted to *2-3%*. In our experiments, we were unfortunately unable to directly measure the increase of the critical temperature $T_C$ in zirconium during the process of low temperature straining without the interim heating up to the room temperature. It is clear that the effect of increase of the zirconium critical temperature $T_C$ without the warm up to the room temperature will be much bigger and even comparable, for instance, with the effect, observed in rhenium [10], which also has the *hcp* crystal lattice.

As shown in [11], the deformation of the same purity zirconium at the temperature of *4.2 K* is accompanied by an appearance of a number of the point, linear and planar defects; and the process of heating of researched samples up to the room temperature restores about *60 %* of the magnitude of electrical resistance as a result of the annihilation, dissociation and removal of certain types of point defects and point defects clusters.

The research on the zirconium structure after its deformation by the extension of *8 %* at the temperature of *4.2 K*, using the electron microscopy, shows that the deformation develops via the slippage and twinning, and the later one is the predominant physical mechanism. The structure of zirconium exhibits the comparatively large twins as well as the stacks of parallel micro-twins as shown in Fig. 1. The accommodation zones in close proximity to the twin layer boundaries in zirconium are, as a rule, less pronounced, comparing to the niobium [6]. In the areas far from the twins, where the plastic deformation develops via the slippage, the average dislocation density amounts to the value of ~$10^{10}cm^{-2}$.

The substantial increase of the magnitude of the critical temperature $T_C$, observed in the zirconium, is in agreement with the earlier formulated considerations [12] on the relation between the change of the critical temperature $T_C$ and the electron attraction constant *g* in the *Barden Cooper Schrieffer* (*BCS*) theory [13]. According to this theory, the critical temperature $T_C$, *Debye* temperature *θ*, and electron attraction constant *g* are interrelated as

$$T_C = 1.14\theta \exp(-1/g).$$

Taking to the account that $g = N(0)U$, where *N(0)* is the density of electron states at the *Fermi* surface and *U* is the electron – lattice coupling constant, we can determine the variations of the electron attraction constant *g* by considering the changes of both *N(0)* and *U*. The physical mechanism of the critical temperature shift was researched earlier, for instance: *1)* by the introduction of the "donor" or "acceptor" impurities into the metal [14] or *2)* by the influence of the hydrostatic pressure on the superconducting properties [15]. However, in the case of the deformation application, the appearing crystal lattice defects give rise to the new local and quasi-local crystal vibration modes [16], which can cause the constant *U* to change [17]. Under the appropriate conditions, the defects of a certain type such as the dislocations can make some changes in the value of *N(0)* [18]. In the case of tight-binding superconductors with the big magnitude of the electron

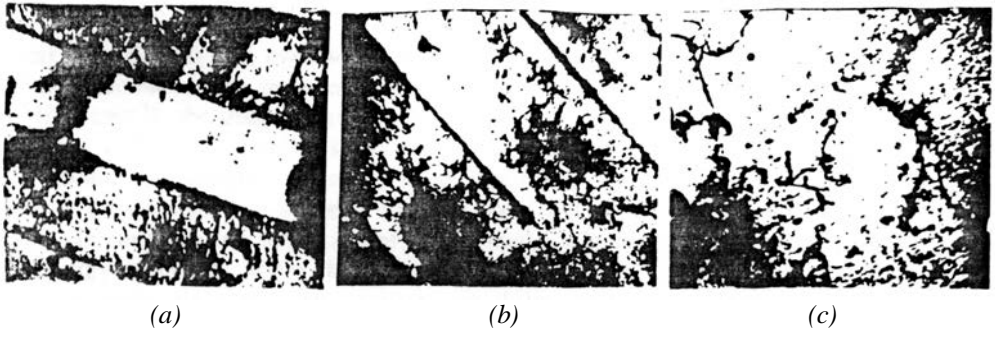

*(a)*           *(b)*           *(c)*

***Fig. 1.** Typical defects in zirconium crystal lattice after its straining by extension of 8 % at temperature of 4.2 K.
(a) Twins (x35,000); (b) Microtwins (x14,000); (c) uniformly distributed dislocations (x35,000)*

attraction constant $g \simeq 0.4$ (lead, mercury), the further increase of this constant is restricted by the crystal lattice stability [19]. In the weak binding superconductors with the small electron attraction constant $g$, the influence by the plastic deformation effect on the critical temperature $T_C$ will be stronger. There are the increased ratios $\Delta T_C/T_{C0}$ in such metals as *Zr*, *Re*, *Sn*, *Nb* for which the electron attraction constants $g$ are *0.15*, *0.18*, *0.25*, *0.30* respectively [20].

The presently available theoretical models provide a possible insight on the role by the various types of *defects* and stresses on the critical temperature shift $T_C$. Taking into the consideration the fact that the dislocations and twin boundaries are the predominant defect types that appear in the zirconium crystal lattice after its low temperature deformation and subsequent heating up to the temperature of *300 K*, let us evaluate the possible effect by these defects and by the relevant internal stress fields on the critical temperature $T_C$.

Let us suppose that, in the experiments with the zirconium, the increase of $T_C$ is brought about by the internal stresses in the crystal lattice. The level of these internal stresses can be estimated, assuming that their action is similar to that of the hydrostatic pressure. Then, knowing the positive value $dT_C/dP \simeq 1.4 \cdot 10^{-4}$ K/MPa in [21], we find that these stresses must be compressive and amount to $\sim 7$ MPa. Thus, the estimated stress is considerably higher than the yield stress of zirconium [22] and, hence, it can essentially be relaxed. Therefore, there is no reason to believe that a presence of stresses of described intensity can be a main source of the critical temperature increase. It is also not reasonable to think that these stresses, should they exist, could be homogeneous, resulting in a shift of the full superconducting state transition curve to the higher temperatures as it is observed experimentally. At the same time, the inhomogeneous elastic stresses of lower level, which relate to the deformation localization areas, are likely responsible for the smearing of the transition curve.

As shown in [23], the increase of $T_C$ in the niobium is due to the electron attraction enhancement, which is facilitated by the dislocation vibrations within the valleys of the *Peierls* potential relief, in accordance with the *Zaitsev* theory, has the form

$$\Delta T_C^d \simeq \frac{\pi}{8}\left(\frac{c_t}{c_l}\right)^4 \frac{E_F m s}{G\hbar} N_d, \quad (1)$$

where $E_F$ is the *Fermi* energy, $m$ is the electron mass at the *Fermi* surface, $G$ is the shear modulus, $\hbar$ is the *Plank* constant, $c_t$ and $c_l$ are the transverse and longitudinal sound velocities, $s$ is the effective sound velocity, a combination of the $c_t$ and $c_l$ in [23]. The equation (*1*) [23] has been derived with the use of expansion in a small parameter (proportional to $T_{c0}^{-1}$) of one of the terms that describe $\Delta T_c^d$ within the framework of the theory in [24]. In view of the small value of $T_{c0}$ in Zr, the approximation, that is allowable for Nb, is generally invalid, so that the eq. (*1*) can be employed only for an approximate estimation of the effect. To estimate the $\Delta T_c^d$ in such a way, we use the approximation: $m \sim m^*$ ($m^*$ is the cyclotron mass; in the case of zirconium: $m^* \sim 1.5 \cdot m_0$ [25], where $m_0$ is the electron mass) and put $c_t/c_l \simeq 0.67$ (which is typical for metals). By substituting these values of $m$ and $c_t/c_l$ in eq. (*1*) as well as the known values $s \simeq 5 \cdot 10^5$ cm/sec in [26]; $G = 3.27 \cdot 10^{10}$ Pa in [24]; $E_F \sim 7.9$ eV in [27], we finally obtain

$$\Delta T_c^d \sim 8 \cdot 10^{-14} N_d [K]. \quad (2)$$

It is clear from eq. (*2*) that, even within the plastic strain localization areas (the accommodation zones in close proximity to the twin boundaries, where $N_d \sim 10^{11}$ cm$^{-2}$), the dislocation related increase of the critical temperature $T_c$ will amount to $\sim 10^{-2}$ K. This value is less than the experimental $\Delta T_c$. It should be, however, kept in mind that the above estimation is of qualitative character. We also point out to the following circumstance. As shown in [18], in the metals with the special energy spectrum, where the *Fermi* level falls on the region with the discrete electron levels, localized on the dislocations, the variation of the pressure at the low temperatures originates the oscillations of the thermodynamic potential. As a result, a number of the thermodynamic quantities also oscillate. The oscillations of the critical temperature $T_c$ were predicted within the framework of such an approach in [28]. Therefore, in agreement with the results in [28], the certain correction to the magnitude of the dislocation induced critical temperature increase is not excluded in the case of the zirconium with the dislocations at the stress of the entire ensemble of other defects. Thus, the question of a more consistent theoretical treatment of the influence by the dislocations on the critical temperature $T_C$ in the zirconium remains open.

The increase of the critical temperature $T_c$ can also be a consequence of the localized superconductivity, associated with the twin boundaries. The possible physical mechanisms of such superconductivity are considered in the review [20]. The exact information about the fine structure of the real twin boundaries is needed to quantitatively compare the experimental results with the predictions of the localized superconductivity models in the case of zirconium and other superconductors as well.

Thus, in this research paper, we cannot give our preference to any particular physical mechanism of the critical temperature increase at the *N-S* transition in the zirconium. In our opinion, the significant research progress in the considered research field would be achieved, if the materials with the single type of defects could be researched.

## Conclusion

We would like to note that the research results, presented in this research paper and in our previous publications, provide the multiple evidences that the registered increase of critical temperature $T_C$ of zirconium, which occurs due to the low temperature plastic deformation, is typical for the pure metallic superconductors irrespective of their crystal lattice type. It is possible to suppose that the lower the ratio of the inter-electron attraction constant to the maximum value of this parameter, which is the characteristic of the tight bounding superconductors ($g \simeq 0.4$), the stronger the phonon spectrum softening and the bigger the electron-phonon coupling enhancement in the plastic deformation localization areas in the superconductor.

This experimental research is completed in the frames of the fundamental and applied superconductivity research program at *the Schubnikov Cryogenic Laboratory at the National Scientific Centre Kharkov Institute of Physics and Technology* (*NSC KIPT*) in Kharkov in Ukraine. This experimental research program is funded by *the Ukrainian State Committee for the Science and Technologies*.

This research paper was published in the *Low Temperature Physics* (*FNT*) in 1993 in [29].

[*]E-mail: ledenyov@kipt.kharkov.ua

———————